\documentclass[pra, 10pt, twocolumn, letter, floatfix]{revtex4}

\usepackage{amsmath}
\usepackage{amsfonts}
\usepackage{amssymb}
\usepackage{graphicx}
\usepackage{rotating}
\allowdisplaybreaks
\usepackage{setspace}
\usepackage{bm}

\begin{document}

\title{An efficient variational principle for the direct optimization of excited states}

\author{Luning Zhao$^{1,2}$ and Eric Neuscamman$^{1,2,}$\footnote[1]{Electronic mail: eneuscamman@berkeley.edu}}
\affiliation{$^1$Department of Chemistry, University of California, Berkeley, California 94720, USA\\
             $^2$Chemical Sciences Division, Lawrence Berkeley National Laboratory, Berkeley, California 94720, USA}

\begin{abstract}
We present a variational function that targets excited states directly based on their position in the energy spectrum,
along with a Monte Carlo method for its evaluation and minimization whose cost scales polynomially
for a wide class of approximate wave functions.
Being compatible with both real and Fock space and open and periodic boundary
conditions, the method has the potential to impact many areas of chemistry, physics, and materials science.
Initial tests on doubly excited states show that using this method, the Hilbert space Jastrow antisymmetric geminal power ansatz
can deliver order-of-magnitude improvements in accuracy relative to equation of motion coupled cluster theory,
while a very modest real space multi-Slater Jastrow expansion can achieve accuracies within 0.1 eV of the best theoretical benchmarks
for the carbon dimer.
\end{abstract}

\maketitle



The ground state variational principle is probably the most important technique in modern electronic structure theory.
Through its roles in optimizing Slater determinants in Hartree Fock (HF) \cite{Szabo-Ostland} and density functional theory
(DFT) \cite{Parr-Yang}, the matrix product state (MPS) in density matrix renormalization group (DMRG)
\cite{Schollwock:2005:dmrg_review,Chan:2011:dmrg_in_chem},
trial functions in variational Monte Carlo (VMC) \cite{FouMitNeeRaj-RMP-01}, and linear combinations in configuration
interaction (CI) \cite{Helgaker_book}, it exists as a critical element in the vast majority of ground state electronic
structure methods used today.
Its success rests on the existence of a function
\begin{align}
E(\Psi) = \frac{\langle\Psi|H|\Psi\rangle}{\langle\Psi|\Psi\rangle}
\label{eqn:gs_var_func}
\end{align}
whose global minimum is the Hamiltonian's ground eigenstate.
This function provides a metric telling us which parameterization of an approximate ansatz is closest
to the true ground state, thus allowing us to optimize the ansatz's full variational freedom for that state alone without
regard to the description of any other state.
In practice, of course, we are constrained in our choice of ansatz to those permitting an efficient evaluation of $E$.
This constraint notwithstanding, the ground state variational principle has become an essential part of most electronic
structure methods, even those like coupled cluster (CC) theory \cite{BARTLETT:2007:cc_review} whose practical application
involves non-variational methods as well.

To date, the lack of an efficient analogous function for excited states has hindered the development of methods
that can target such states in the same individual and variational way.
Instead, existing excited state methods typically require an ansatz to use its variational freedom to satisfy the needs of many
eigenstates simultaneously, the difficulty of which has limited our predictive power over the doubly-excited states
in light harvesting systems, the spectra of excited state absorption experiments, and the band gaps of transition metal oxides.
For example, linear response (LR) methods such as time dependent HF and DFT \cite{HeadGordon:2005:cis_review},
CI singles (CIS) \cite{HeadGordon:2005:cis_review}, equation of motion CC with singles and doubles (EOM-CCSD) \cite{Krylov:2008:eom_cc_review},
and LR DMRG \cite{Chan:2014:lrdmrg,Chan:2013:thouless,Verstraete:2013:mps_tangent}
are limited by the requirement that all excited states of interest must be found in the ground state's LR space,
which for a nonlinear ansatz is typically much less flexible than its full variational space.
In many other cases, such as state-averaged complete active space methods \cite{Werner:1985:mcscf,Finley:1998:multistate_caspt2},
some VMC approaches \cite{Filippi:2009:sa_vmc},
and directly targeted DMRG \cite{Dorando:2007:hd_dmrg},
crucial ansatz components (often the one particle basis) are required to be the same for the ground and all excited states.
While these methods clearly do not take full advantage of an ansatz's variational freedom,
they have been preferred due to the lack of an efficiently optimizable function whose minimum is an excited state.

This report presents a new variational principle consisting of two parts:  first, a function $\Omega(\Psi)$ whose global minimum is an
excited eigenstate, and second, a method for evaluating and minimizing $\Omega$ whose cost scales polynomially for a wide class of
approximate wave functions.
We will begin by proving that $\Omega$ has the necessary properties to be the basis of an excited state variation principle,
after which we will detail our method for minimizing it.
During this discussion, we will explain which wave functions are compatible with the approach, as well as its general applicability in
chemistry, physics, and materials science.
Finally, we will present numerical examples that demonstrate the method's potential.

We employ the function
\begin{align}
\Omega(\Psi) = \frac{\langle\Psi|(\omega-H)|\Psi\rangle}{\langle\Psi|(\omega-H)(\omega-H)|\Psi\rangle}
\label{eqn:es_var_func}
\end{align}
where the energy shift $\omega$ is assumed to be placed in between distinct eigenvalues of H in order
to target the eigenstate whose eigenvalue is immediately above $\omega$ in the ordered eigenvalue spectrum.
Assuming real numbers for brevity, we proceed to prove that this eigenstate is the global minimum of $\Omega$
as follows.
First, we write an exact ansatz as a linear combination of all eigenstates of $H$, $|\Psi_e\rangle = \sum_i c_i |i\rangle$,
and rewrite $\Omega$ in terms of $H$'s eigenvalues.
\begin{align}
\Omega(\hspace{0.1mm}\vec{c}\hspace{0.3mm}) = \frac{\sum_i c_i^2 (\omega - E_i)}{\sum_i c_i^2 (\omega - E_i)^2}
\end{align}
Differentiating with respect to the elements of $\vec{c}$, we see that $\vec{c}$ is a stationary point (SP) if and only if
\begin{align}
0 = c_i (\omega-E_i) \big( 1 - (\omega-E_i) \Omega \big) \quad \forall \quad i.
\end{align}
Recalling that $\omega$ is assumed to be distinct from any of $H$'s eigenvalues, we see that $\vec{c}$ cannot
be a SP if any two of its elements that correspond to distinct
Hamiltonian eigenvalues are non-zero, as this would prevent $(1 - (\omega-E_i) \Omega)$ from vanishing for both of them.
In other words, $|\Psi_e\rangle$ cannot be a SP of $\Omega$ unless the nonzero
values in $\vec{c}$ all correspond to one (possibly degenerate) eigenvalue of $H$.
As the eigenstates of $H$ are clearly SPs of $\Omega(\vec{c})$,
we see that $|\Psi_e\rangle$ is a SP if and only if it is an eigenstate.
We thus conclude that the global minimum of $\Omega(\vec{c})$ is the eigenstate whose eigenvalue is immediately
above $\omega$, or a linear combination of such eigenstates if this eigenvalue is degenerate.
As $|\Psi_e\rangle$ can describe any state in Hilbert space, this minimum value will be less than or equal to that
of any approximate ansatz, thus achieving the variational property we desire.

While formally interesting, the mere existence of a variational function for excited states is not useful without
an efficient way to evaluate and minimize it.
Indeed, the presence of $H^2$ makes the straightforward evaluation of $\Omega$ drastically more
expensive than the ground state function $E$, which is why studies that have worked implicitly with this function in
the past \cite{Dorando:2007:hd_dmrg,Carter:2011:hd_ci} have, to the best of our knowledge, always approximated this
term (see discussion of harmonic Ritz methods below).
Here we avoid explicitly squaring $H$ by resolving identities via complete sums over states,
\begin{align}
\Omega(\Psi) = \frac{\sum_m \langle\Psi|m\rangle\langle m|(\omega-H)|\Psi\rangle}
                    {\sum_m \langle\Psi|(\omega-H)|m\rangle\langle m|(\omega-H)|\Psi\rangle}
\label{eqn:ri_in_omega}
\end{align}
which we may evaluate (up to a controllable statistical uncertainty that obeys the zero variance principle)
through Monte Carlo integration as
\begin{align}
\Omega_{MC}(\Psi) = \frac{\sum_{m\in \xi} W_m}{\sum_{m\in \xi} W_m^2} \qquad W_m \equiv \frac{\langle m|(\omega-H)|\Psi\rangle}{\langle m|\Psi\rangle}
\end{align}
where the elements of $\xi$ are sampled from $|\langle m|\Psi\rangle|^2$ via a Metropolis walk
(note that the normalization constants for numerator and denominator cancel).
Thus any ansatz admitting efficient evaluations for $W_m$ will be compatible with our approach.
This includes the wide class of wave functions already used in ground state VMC, such as
Slater-Jastrow (SJ) \cite{FouMitNeeRaj-RMP-01},
multi-Slater-Jastrow (MSJ) \cite{Morales:2012:msj},
and the Jastrow antisymmetric geminal power (JAGP)
\cite{Sorella:2003:agp_sr,Sorella:2004:agp_sr,Sorella:2007:jagp_vdw,Sorella:2009:jagp_molec,Neuscamman:2012:sc_jagp,Neuscamman:2013:jagp}
as well as the MPS.
Moreover, the method is applicable to real space (in which case $m$ is a position vector) and Fock space
(in which case $m$ is an occupation number vector), in open systems (e.g.\ chemistry)
or period ones (e.g.\ materials' band gaps),
thus enabling essentially all of the tools we have for ground state VMC to be employed
in the direct optimization of an individual, $\omega$-targeted excited state.

To this purpose, we now present a minimization method analogous to the linear method (LM)
\cite{UmrTouFilSorHen-PRL-07,TouUmr-JCP-08}
used for ground states.
Replacing the original ansatz with a linear combination of itself and its first derivatives with respect
to its parameters $\vec{u}$,
\begin{align}
\label{eqn:deriv_expansion}
& |\Psi\rangle \rightarrow \sum_i a_i |\Psi^i\rangle, \\
\label{eqn:deriv_def}
& |\Psi^0\rangle \equiv |\Psi\rangle \qquad |\Psi^i\rangle \equiv \partial/\partial u_i|\Psi\rangle \hspace{4mm} \forall \hspace{4mm} i>0,
\end{align}
we may then minimize $\Omega$ with respect to $\vec{a}$ by solving the generalized eigenproblem
\begin{align}
\label{eqn:gen_eigenproblem}
\sum_j \langle\Psi^i| \big[ \hspace{1mm} (\omega-H) \hspace{1mm} - \hspace{1mm} \lambda \hspace{1mm} (\omega-H)^2 \hspace{1mm} \big] |\Psi^j\rangle a_j = 0.
\end{align}
Assuming we were already near the minimum, in which case all $\vec{a}$ elements except $a_0$ will be small,
then we may use $\vec{a}$ to update $\vec{u}$ through a reverse Taylor expansion exactly as in the
traditional LM (in practice we may also shift the eigenproblem as in the LM to ensure this assumption is valid).
This entire procedure is then iterated in similar fashion to Newton's method until the minimum of $\Omega$ is reached,
thus optimizing both the linear and nonlinear parameters of our original ansatz $|\Psi\rangle$.
As the matrix elements for the eigenproblem can be evaluated through the same stochastic identity resolution
as described above, we arrive at a full-fledged and efficient method for the evaluation and minimization of $\Omega$ for
any ansatz that can be efficiently used with the ground state LM.
The precise polynomial cost scaling will of course depend on the choice of $\Psi$, with examples including
$N_s N_e^3$ for real space SJ and JAGP and $N_s N_e^4$ for Hilbert space JAGP, where $N_s$ and $N_e$ are the number
of samples and electrons, respectively, both of which will grow linearly with system size.

Note the similarity of this eigenvalue equation to the harmonic Davidson equation that arises in applications
\cite{Tackett:2002:targeting_eigenvecs,Dorando:2007:hd_dmrg,Carter:2011:hd_ci,Kresse:2012:interior_eigensolver}
of the harmonic Ritz principle
\cite{Morgan:1991:hd_ci,Sleijpen:1996:hjd}
for targeting interior eigenvalues of a matrix.
In fact, some of these approaches
\cite{Dorando:2007:hd_dmrg,Carter:2011:hd_ci}
appear to have been minimizing an approximation to $\Omega$
with respect to linear parameters, in which $PH^2P$ was approximated by $PHPHP$, where $P$ is the projector
into the subspace corresponding to the linear parameters in question.
Except for its controllable statistical uncertainty, the present approach makes no approximation when evaluating $\Omega$
and can optimize both linear and nonlinear parameters alike.

\begin{figure}[b]
\includegraphics[width=8.5cm]{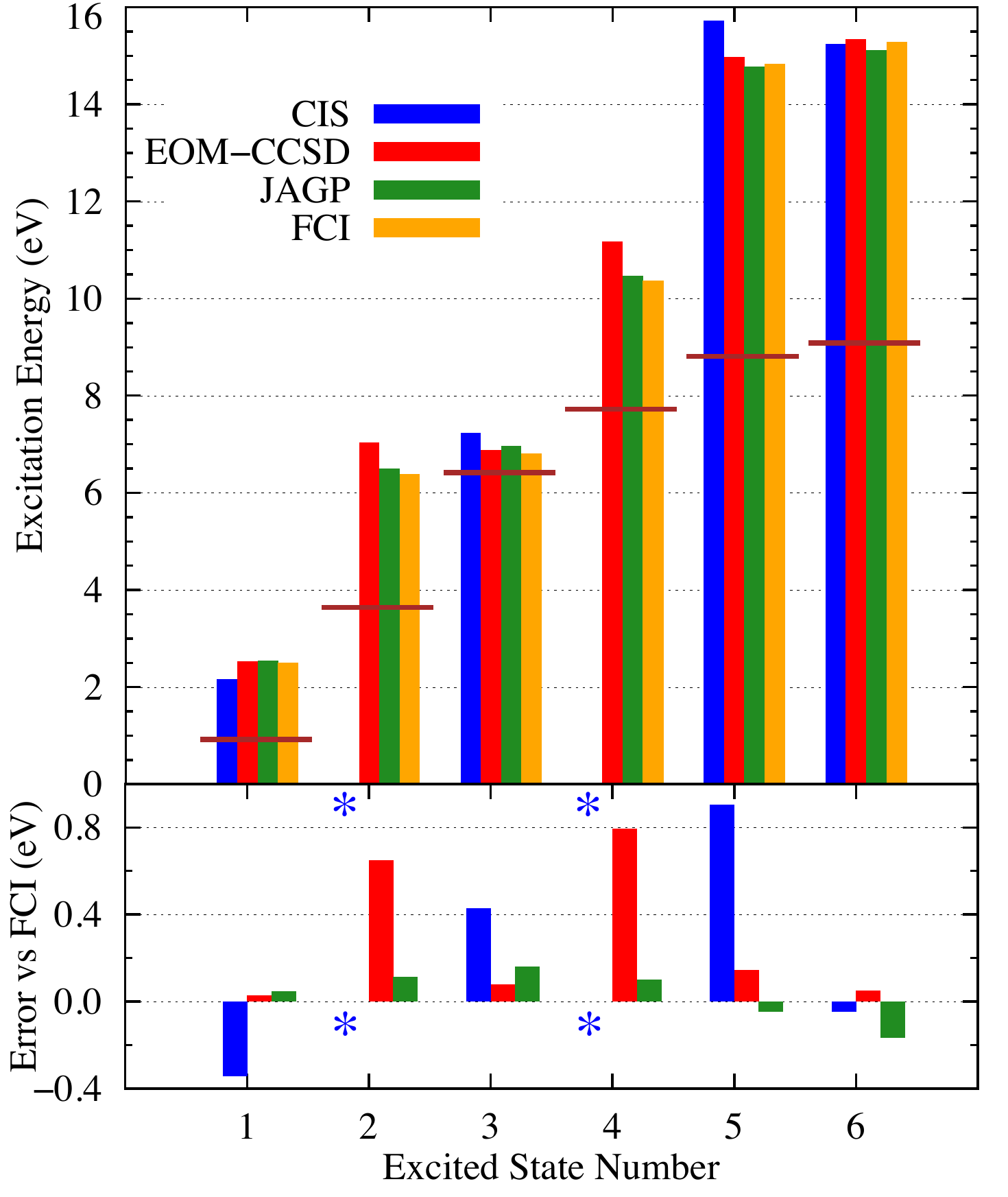}
\caption{Singlet excitations for CH$_2$ in a STO-3G basis.
         Lines mark $\omega$ values.
         Asterisks mark doubly excited states.
        }
\label{fig:ch2}
\end{figure}

\begin{figure}[b]
\includegraphics[width=8.5cm]{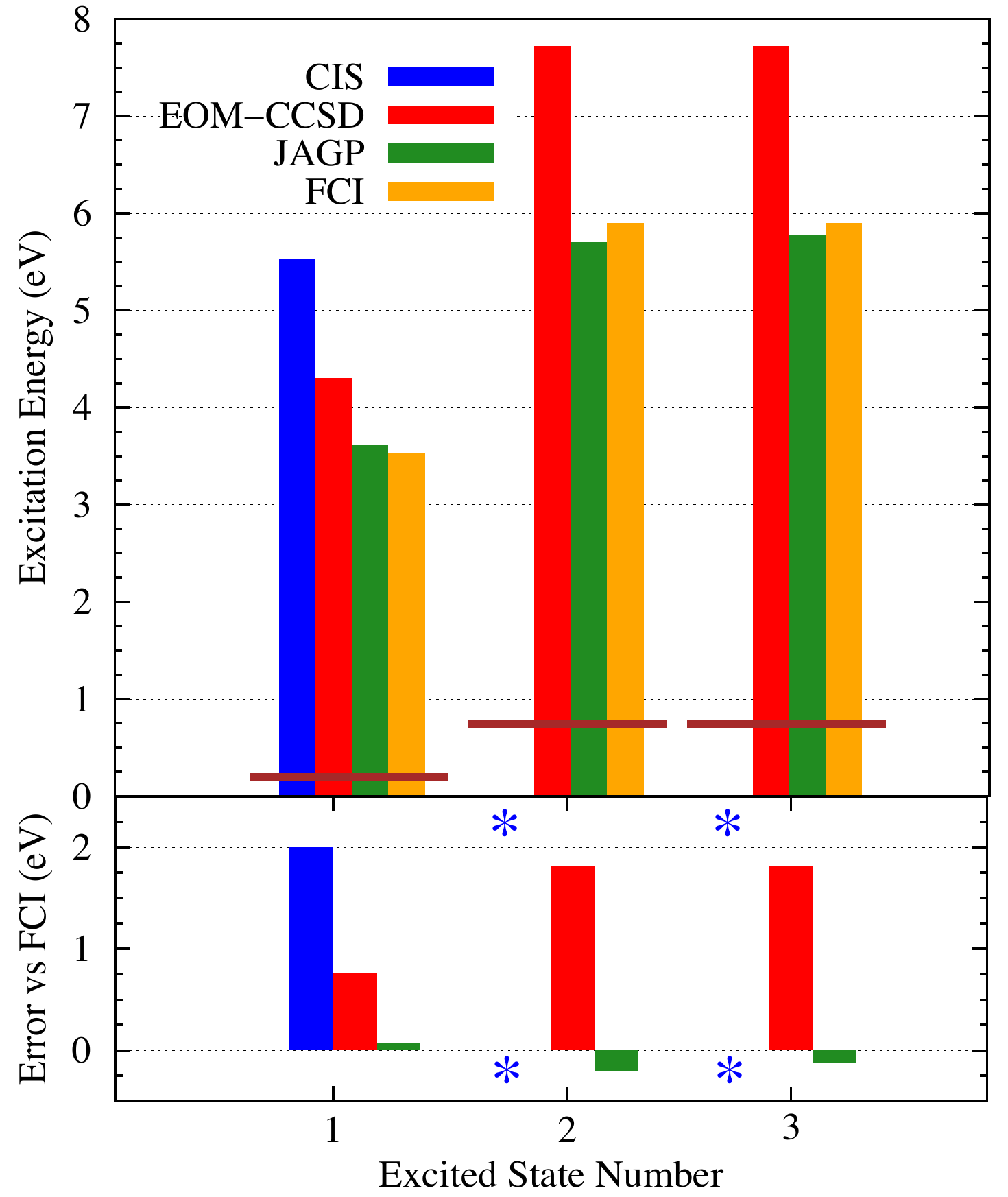}
\caption{Singlet excitations for H$_6$ in a 6-31G basis.
         Lines mark $\omega$ values.
         Asterisks mark doubly excited states.
        }
\label{fig:h6}
\end{figure}

Before discussing results, we should consider the formal consequences of employing an approximate ansatz.
First, the excitation energies we report are taken as the difference in $E(\Psi)$ between the ground and excited states.
One could instead compute energies as $E_\Omega=\omega-\Omega^{-1}$ on the assumption that each $\Psi$ was an exact eigenstate,
but we find this approximation to be much less accurate than computing $E(\Psi)$ directly.
Second, due to the nonzero energy variance of an approximate ansatz, the $\omega$ value at which $\Omega$'s global minimum switches states
tends to lie lower in energy than it would for an exact ansatz, a trend that all of our results display.
Third, as occurs for $E(\Psi)$ in the ground state, the global minimum of $\Omega(\Psi)$ for an approximate ansatz may break symmetry.
Fourth, the SPs of $\Omega$ will not necessarily be orthogonal to one another, either within those for a single shift $\omega$ or between different shifts,
but they may be orthogonalized after the fact by diagonalizing the Schr\"{o}dinger equation within the basis they define.
In the present results, doing so was only necessary for CH$_2$'s 5th and 6th excited states,
which our method found as symmetry-broken mixtures of one another and for which we report the post-2-state-diagonalization energies.
While these various complications are undesirable, one should remember that
the ground state function $E(\Psi)$ has been enormously successful despite suffering the same deficiencies.
Indeed, the whole point of the variational principle is that as $\Psi$ becomes more flexible these issues must abate, and so just as in the ground state case,
our excited state variational principle offers a systematically improvable path towards resolving its own difficulties.

As a demonstration in Fock space, we have used the method to optimize the Hilbert space JAGP ansatz \cite{Neuscamman:2013:jagp} for singlet excited states in
CH$_2$ (Figure \ref{fig:ch2}), an H$_6$ ring (Figure \ref{fig:h6}), and C$_2$ (Figure \ref{fig:c2_631g}).
(Full computational details for all demonstrations are available in the appendix.)
In CH$_2$, the two doubly excited states are absent in CIS due to HF's limited LR space and are treated poorly by EOM-CCSD.
While CCSD's LR space contains doubles, it lacks the triples necessary to describe the
orbital relaxations that should accompany the excitation.
Although JAGP's LR space also lacks triples, it agrees much better with full CI (FCI) \cite{Helgaker_book}, because the variational
minimization of $\Omega$ explores regions of parameter space beyond the LR regime.
The excited states of H$_6$ were more complicated, each having FCI expansions with 12 or more determinants with normalized coefficients above 0.1 as compared
to 8 or less for CH$_2$.
Nonetheless, the same pattern emerges:  the large errors in EOM-CCSD are reduced by an order of magnitude in variationally optimized JAGP.

C$_2$ provides further evidence of JAGP's superiority to EOM-CCSD for double excitations while also revealing the limits
of the ansatz's flexibility.
While JAGP delivers 0.1 eV accuracy versus FCI for excited states 1, 2, 4, and 5, it shows an error almost as large
as EOM-CCSD for state 3, a complicated excitation involving four different electrons in a mixture of double excitations.
Even with this difficulty, which arises due to the inherently two-electron nature of the ansatz, JAGP proves more
accurate than EOM-CCSD for each of the first five singlet excitations of C$_2$.


\begin{figure}[b]
\includegraphics[width=8.5cm]{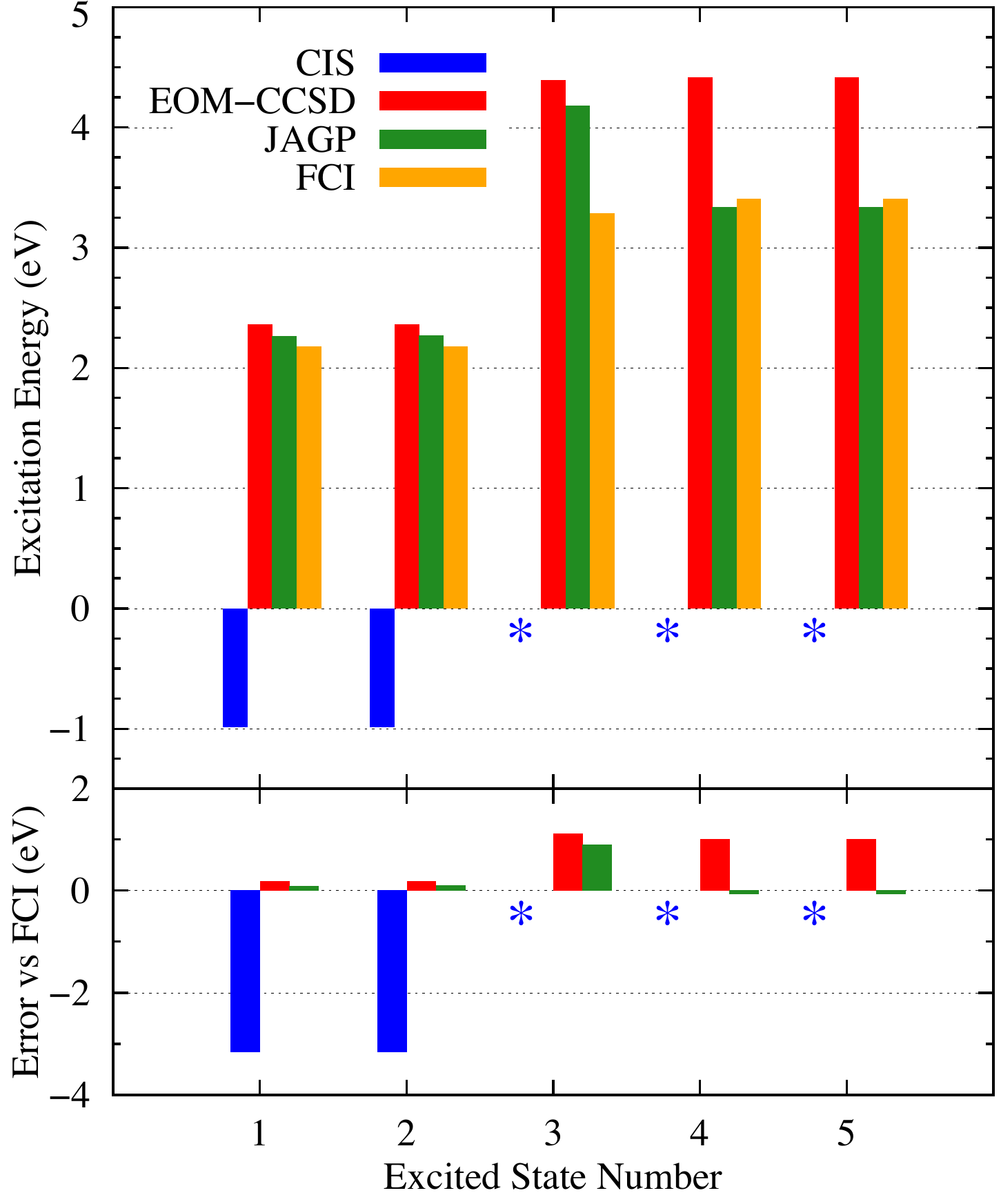}
\caption{Singlet excitations for C$_2$ in a 6-31G basis.
         Asterisks mark doubly excited states.
        }
\label{fig:c2_631g}
\end{figure}

\begin{figure}[b]
\includegraphics[width=8.5cm]{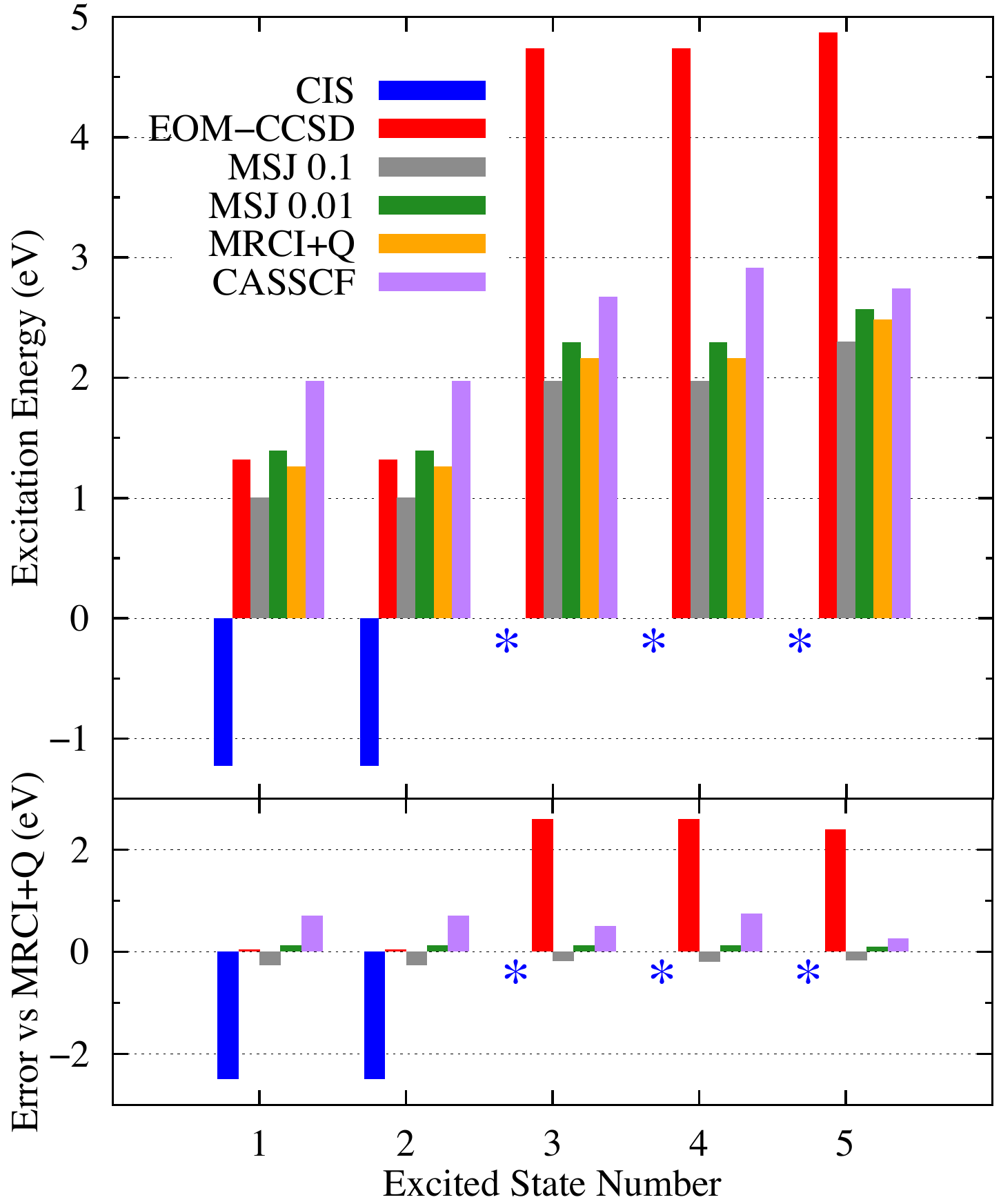}
\caption{Singlet excitations for C$_2$ in a cc-pVTZ basis with MSJ in real space.
         Asterisks mark doubly excited states.
        }
\label{fig:c2_cc-pVTZ}
\end{figure}

To showcase the method's systematic improvability and compatibility with a real space Monte Carlo walk, we have also treated C$_2$ with a MSJ ansatz consisting of short configuration
state function (CSF) expansions and spline-based 1- and 2-body Jastrow factors (Figure \ref{fig:c2_cc-pVTZ}).
For each state, we selected CSFs with coefficients above a given threshold from a complete active space (CAS) wave function,
leading to fewer than 10 (65) CSFs per state for a threshold of 0.1 (0.01).
Under variational optimization (with the random walk now in real space), the worst-case MSJ excitation energy error is found to drop from 0.3 to 0.1 eV 
upon lowering the threshold, as expected for a systematically improvable method.
As a benchmark we use Davidson-corrected multi-reference CI (MRCI+Q) in a triple-zeta basis, which for excited state 5 (the $^1\Sigma_g^+$ state) is within 0.03 eV of the recent
quadruple-zeta DMRG \cite{Sharma:2015:non_alb_dmrg} and FCI quantum Monte Carlo (FCIQMC) \cite{Blunt:2015:excited_fciqmc} benchmarks.
Significantly, our MSJ result for this state (2.57eV) is within 0.1 eV of these benchmarks (2.47eV), despite containing fewer than 100 variational parameters,
compared to more than 4,000 in EOM-CCSD, millions in DMRG, and 2,000 in the FCIQMC trial function.
This success, along with MSJ's high accuracy for C$_2$'s other excited states, demonstrates the advantage of optimizing an ansatz directly and variationally for an
individual excited state.


We have presented a Monte Carlo method for the efficient, variational optimization of a function $\Omega$ whose global minimum can be tuned
to target individual excited states and which may be used at polynomial cost with a wide range of approximate wave functions.
In demonstrations on three small molecules with low-lying doubly excited states, the method's ability to explore
an ansatz's full variational freedom allows for drastic improvements in accuracy compared to linear response theories such as
EOM-CCSD, which stands as the current state-of-the-art in polynomial-cost methods for excited states in chemistry.
Further, we have shown that for the notoriously difficult double excitations of the carbon dimer, variational optimization allows a \textit{very} modest
multi-Slater Jastrow expansion to achieve accuracies on par with the much more cumbersome DMRG and FCIQMC benchmarks.
Given the central importance of double excitations in light harvesting and excited state absorption experiments, the
method's compatibility with periodic boundary conditions and thus the solid state, its systematically improvable nature,
and its direct connection to the most widely used method in ground state electronic structure,
we look forward to further exploring its usefulness in modeling challenging excited states.

We acknowledge funding from the Office of Science, Office of Basic Energy Sciences, the US Department of Energy, Contract No. DE-AC02-05CH11231.

\appendix

\section{General Information}

EOM-CCSD and FCI results were computed with \uppercase{MOLPRO} \cite{MOLPRO_brief},
CIS results with QChem \cite{QChem:2006,QChem:2013}, MSJ results with a modified version of QMCPACK \cite{Ceperly:2012:hybrid_alg_qmc,Ceperly:2012:full_accel_qmc}
with the CAS truncation taken from GAMESS \cite{Schmidt:1993:gamess},
and JAGP results with our own prototype Hilbert space
quantum Monte Carlo code with one- and two-electron integrals imported from Psi3 \cite{Psi3}.
In JAGP we worked exclusively in the symmetrically orthogonalized ``$S^{-1/2}$'' one particle basis
and froze the C 1s orbital at the HF level.
All statistical uncertainties were converged to less than 0.01eV in all cases.

\section{CH$_2$}

For CH$_2$ we used a minimal STO-3G basis set \cite{Pople:1969:sto-3g} and shifts in Hartree of $\omega$ = 
-38.4,
-38.3,
-39.198,
-38.15,
-38.110, and
-38.1 for excited states 1 to 6, respectively.
As mentioned in the main text, this resulted in minima of $\Omega$ for the last two shifts that
corresponded JAGPs that were symmetry broken combinations of the 5th and 6th excited states.
The numbers we report for the excitation energies are those after rediagonalizing the Sch\"{o}dinger
equation in the basis of these two JAGP wave functions, which restores the desired symmetry.
Finally, in \AA, the CH$_2$ geometry was

\vphantom{m}
\begin{tabular}{l r r r}
  \hspace{0mm} C & \hspace{2mm} -0.0722376285 & \hspace{2mm}      -0.0574604043 & \hspace{2mm}       0.0000000000 \\
  \hspace{0mm} H & \hspace{2mm} -0.0198102890 & \hspace{2mm}       1.0990427214 & \hspace{2mm}       0.0000000000 \\
  \hspace{0mm} H & \hspace{2mm}  1.0664179823 & \hspace{2mm}      -0.2665333714 & \hspace{2mm}       0.0000000000 \\
\end{tabular}
\vspace{2mm}

\section{H$_6$}

For H$_6$ we used the 6-31G basis \cite{POPLE:1972:6-31g_basis}
and shifts in Hartree of $\omega$ = 
-3.17, -3.15, and -3.15 for excited states 1 to 3, respectively.
The geometry for H$_6$ was chosen as a regular hexagon with edge lengths (i.e.\ bond distances) of 1.5 \AA.

\section{Fock Space C$_2$}

For C$_2$ with a Fock space random walk we used the 6-31G basis \cite{POPLE:1972:6-31g_basis}
and shifts in Hartree of $\omega$ = 
-74.85,
-74.85,
-74.65,
-74.60, and
-74.60
for excited states 1 to 5, respectively.
Note that these shifts were not plotted in the main text as they are all \textit{below} the
JAGP ground state energy of -75.5915 Hartree and would not conveniently fit on the plot.
Recall that the finite variance of an approximate wave function causes the values of
$\omega$ at which the $\Omega$-minimum switches states to shift down in energy, and indeed
in C$_2$ this effect appears to be large enough to push the switching energies below the
ground state energy.
None the less, when $\Omega$ is minimized for these shifts and then the energy $E(\Psi)$
is calculated for the resulting JAGP states, the results are those plotted in the main text.
In future work, we believe it will be possible to build modified versions of $\Omega$ that
share its formal properties while also supressing this ``early switching'',
but for now we simply choose shifts based on where the $\Omega$-minimum switches states.
The C$_2$ bond distance was 1.2425146399 \AA.

\section{Real Space C$_2$}

For C$_2$ with a real space random walk we used the cc-pVTZ basis \cite{Dunning:1993:basis_boron_neon}, both for the
orbitals of the MSJ wave function and for the CIS, EOM-CCSD, CASSCF, and MRCI+Q calculations.
The CAS expansion from which CSFs were were taken for MSJ was the CAS space resulting from an equal weighted state averaged
(8e,8o) CASSCF calculation including the ground state and the first 5 singlet excited states.
The CSF orbitals were taken as the optimized CASSCF orbitals.
The Jastrow factors (one each for electron-nuclear, opposite-spin-electron, and same-spin-electron) were one
dimensional functions of the magnitude of the interpartical distance, the natural logarithms of which were
parameterized as a 10-section bspline with a cutoff radius of 5 Bohr for the electron nuclear and 10 Bohr
for the electron-electron.

We found that the downshifting of the switching values of $\omega$ was even more
severe in real space, and that a significant difference was seen when finding the
ground state by minimizing $E(\Psi)$ versus $\Omega(\Psi)$ (note that no significant
difference of this type was seen in the Fock space cases).
For consistency, we minimized $\Omega(\Psi)$ for all states, including the ground
state, whose shift $\omega$ was chosen to lie near the point at which the minimum
switched to the first excited state.
We observed that so long as it was near the switching point, the precise choice of
$\omega$ in the ground state optimization had only a very minor effect on
the predicted excitation energies, whereas optimizing $E(\Psi)$ instead (equivalent
to choosing $\omega=-\infty$) produced a substantial ground state bias.
The $\omega$ values used for the reported excitation energy calculations were
-79.15 for the ground state and
-79.00,
-79.00,
-78.70,
-78.70, and
-78.68
for excited states 1 to 5, respectively.
The C$_2$ bond distance was 1.2425146399 \AA.

\bibliographystyle{aip}
\bibliography{direct_excited_vmc}

\end{document}